# Exfoliation and van der Waals heterostructure assembly of intercalated ferromagnet $Cr_{1/3}TaS_2$

Yuji Yamasaki[1], Rai Moriya*[,1], Miho Arai[1], Satoru Masubuchi[1], Sunseng Pyon[2], Tsuyoshi Tamegai[2], Keiji Ueno[3], and Tomoki Machida*[,1]

[1] *Institute of Industrial Science, The University of Tokyo, 4-6-1 Komaba, Meguro-ku, Tokyo 153-8505, Japan*

[2] *Department of Applied Physics, The University of Tokyo, 7-3-1 Hongo, Bunkyo-ku, Tokyo 113-8656, Japan*

[3] *Department of Chemistry, Graduate School of Science and Engineering, Saitama University, 255 Shimo-Okubo, Sakura-ku, Saitama City, Saitama 338-8570, Japan*

*E-mail: moriyar@iis.u-tokyo.ac.jp; tmachida@iis.u-tokyo.ac.jp

**Abstract.** Ferromagnetic van der Waals (vdW) materials are in demand for spintronic devices with all-two-dimensional-materials heterostructures. Here, we demonstrate mechanical exfoliation of magnetic-atom-intercalated transition metal dichalcogenide $Cr_{1/3}TaS_2$ from its bulk crystal; previously such intercalated materials were thought difficult to exfoliate. Magnetotransport in exfoliated tens-of-nanometres-thick flakes revealed ferromagnetic ordering below its Curie temperature $T_C \sim 110$ K as well as strong in-plane magnetic anisotropy; these are identical to its bulk properties. Further, van der Waals heterostructure assembly of $Cr_{1/3}TaS_2$ with another intercalated ferromagnet $Fe_{1/4}TaS_2$ is demonstrated using a dry-transfer method. The fabricated heterojunction composed of $Cr_{1/3}TaS_2$ and $Fe_{1/4}TaS_2$ with a native $Ta_2O_5$ oxide tunnel barrier in between exhibits tunnel magnetoresistance (TMR), revealing possible spin injection and detection with these exfoliatable ferromagnetic materials through the vdW junction.

1. **Introduction**

Magnetic-atom-intercalated transition metal chalcogenides (TMDs) such as $X$TaS$_2$ and $X$NbS$_2$ ($X$ = Mn, Co, Cr, or Fe) exhibit unique magnetic properties. Owing to their layered crystal structures and large spin-orbit coupling, intercalated TMDs show extraordinarily large magnetic anisotropy [1, 2], giant anisotropic magnetoresistance (AMR) [3], and chiral magnetic soliton [4-6]. Recent successful preparation of graphene, other two-dimensional (2D) materials, and their van der Waals (vdW) heterostructures, inspired the use of intercalated TMD materials as candidates for ferromagnetic 2D materials; such materials can be mechanically exfoliated and transferred onto another 2D material to construct vdW junctions. Demonstrations of a spin injection and detection with such ferromagnetic 2D materials are important building blocks for realizing 2D-material-based spintronics [7-10]. Up to now, several candidates for the 2D-material ferromagnets have been investigated, such as K$_2$CuF$_4$ [11], Fe$_x$TaS$_2$ [3, 12], Cr$_2$Ge$_2$Te$_6$ [13-15], Fe$_3$GeTe$_2$ [16, 17], and CrI$_3$ [18]. Although previously thought to be difficult, mechanical exfoliation of sub-100 nm-thick Fe$_x$TaS$_2$ has recently been reported by different groups [3, 12]. Furthermore, dry-transfer fabrication of a heterojunction between different flakes of Fe$_x$TaS$_2$ and tunnelling magnetoresistance (TMR) effects in the fabricated junction have been demonstrated [12]. More recently, an intrinsic ferromagnetism has been demonstrated in monolayer CrI$_3$ and bi-layer Cr$_2$Ge$_2$Te$_6$ [19, 20]. All of these materials have a magnetic easy axis along the out-of-plane direction and Fe$_x$TaS$_2$ is an only the ferromagnetic material exhibiting metallic conduction. There is a need to explore other exfoliatable metallic ferromagnets that have different anisotropy. Early studies suggested that intercalation of TaS$_2$ with other magnetic atoms such as Mn, Cr, and Co exhibits an in-plane magnetic easy axis [21, 22], although its detailed magnetotransport properties and mechanical exfoliation were not investigated. In this study, we demonstrate mechanical exfoliation and vdW heterostructure assembly of ferromagnetic Cr$_{1/3}$TaS$_2$ together with Fe$_{1/4}$TaS$_2$ and investigate their magnetotransport properties.

The crystal structures of $Cr_{1/3}TaS_2$ and $Fe_{1/4}TaS_2$ are presented in Figs. 1(a) and (b), respectively. In both materials, magnetic atoms are intercalated in the vdW gap between the 2H-stacked $TaS_2$ layers. The occupation of intercalated $Cr^{3+}$ and $Fe^{2+}$ at these compositions is different, as shown in the figures. The intercalated magnetic atoms form a $\sqrt{3}\times\sqrt{3}$ superstructure in $Cr_{1/3}TaS_2$, whereas $Fe_{1/4}TaS_2$ forms a 2×2 superstructure; the unit cell of these superstructures is depicted by the dashed line. The intercalated magnetic atoms provide local magnetic moments and they are ferromagnetically coupled each other through indirect Ruderman-Kittel-Kasuya-Yosida (RKKY) exchange interaction via conduction carriers within the $TaS_2$ layer. For $Cr_xTaS_2$, the only previous study is for the bulk crystal of $x = 1/3$ in the 1980s, and it is reported to have ferromagnetic ordering below its Curie temperature $T_C \sim 110$ K [21, 22]. In the case of the $Fe_xTaS_2$ bulk crystal, ferromagnetic ordering is reported in the range of $0.15 < x < 0.45$ and the highest $T_C \sim 150$ K is obtained at $x = 1/4$ [23, 24].

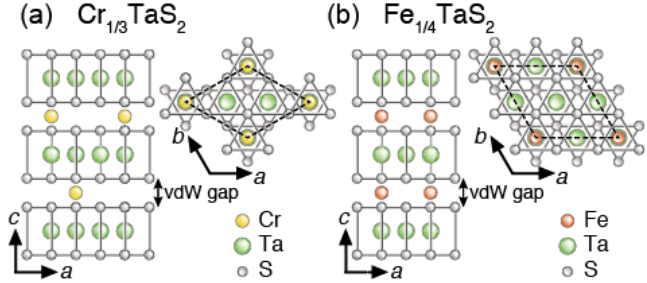

**Figure 1** (a,b) Crystal structures of (a) $Cr_{1/3}TaS_2$ and (b) $Fe_{1/4}TaS_2$, showing side and top views. The dashed line indicates the $\sqrt{3}\times\sqrt{3}$ and 2×2 superstructures of $Cr^{3+}$ and $Fe^{2+}$ ions, respectively.

## 2. Methods

The bulk crystals of $Cr_{1/3}TaS_2$ and $Fe_{1/4}TaS_2$ used in this study were fabricated by chemical vapour transport [2, 23-25] (details are provided in supplementary information). Thin flakes of the materials were exfoliated from bulk crystals by mechanical exfoliation using adhesive tape (31B, Nitto Denko), and deposited on either 300 nm $SiO_2$/doped Si substrate or polydimethylsiloxane (PDMS) sheets (PF-X4, Gelpak). Atomic force microscopy (AFM) was carried out in atmosphere to characterize the surface and thickness of the exfoliated flakes. By using dry transfer [26, 27], a flake on the PDMS sheet was transferred onto the other flake located on the $SiO_2$/Si substrate to fabricate a heterojunction between the different flakes. To minimize surface contamination, a freshly cleaved surface between the flakes was contacted to construct the junction. A contact consisting of 45-nm-Au/85-nm-Ti was fabricated on exfoliated flakes or fabricated heterostructure by electron beam (EB) lithography and EB evaporation. The electrical contacts for the bulk crystals are obtained with silver paste. Magnetotransport was measured with a $^4$He cryostat. Differential resistance $dV/dI$ was measured by applying ac current $I_{AC}$ together with dc bias current $I_{DC}$. Magnetization measurements were performed using a commercial SQUID magnetometer (MPMS, Quantum Design Inc.). Cross-sectional scanning transmission electron microscopy (STEM) and energy dispersive X-ray spectroscopy (EDX) was carried out with a 200 kV electron beam.

## 3. Results and Discussion

The magnetization versus applied magnetic field (*M-B*) curves of both bulk crystals were measured in both in-plane and out-of-plane field directions at 10 K and the results are presented in Figs. 2(a) and (b). For $Cr_{1/3}TaS_2$, we observed magnetization saturation at low magnetic fields when the external field was applied in the in-plane direction. On the other hand, saturation was not observed within the measured field range for the out-of-plane field direction. We also note the steeper slope existing around the low-field region of the out-of-plane *M-B* curve. By contrast, $Fe_{1/4}TaS_2$ exhibited the opposite trend, such that it was difficult to saturate the magnetization when external field was applied in the in-plane direction, whereas square hysteresis was observed when the external field was in the out-of-plane direction. These results indicate that the magnetization easy axes of $Cr_{1/3}TaS_2$ and $Fe_{1/4}TaS_2$ are in-plane and out-of-plane, respectively;

easy axes directions are consistent with the previous report [21, 22]. The strong contrast of the magnetic easy axis in these two materials has been understood from the contribution of the orbital angular moment of intercalated magnetic ions, where a non-negligible orbital angular moment of intercalated Fe ions aligns the total magnetic moment along out-of-plane direction [28]. However, such a contribution is negligible for intercalated Cr ions [21]. The out-of-plane *M-B* curve in $Cr_{1/3}TaS_2$ shown in Fig. 2(a) can be attributed to the existence of two different components within the sample that have different saturation fields. One is the component that is saturated at $B \sim 0.1$ T and the magnetization value at this low-field saturation is about 10% of the total magnetization. The other component that is the majority of the total magnetization did not show saturation within our measurement range. The *M-B* curve exhibits a linear increase of magnetization up to the maximum field of ~5 T; such behaviour is typically observed when a magnetic field is applied along the magnetic hard axis. Therefore, this component has larger magnetic anisotropy. From the extrapolation of data, we speculate hard axis saturation field $H_k \sim 12$ T. The smaller anisotropy component that appeared in the low-field region of the *M-B* curve could be due to the slight misorientation of the $Cr_{1/3}TaS_2$ crystal with respect to the external magnetic field, or from the existence of a misoriented domain within the crystal; however, its origin is not clear at this moment. Note that there have been no reports on the out-of-plane *M-B* curve on this material. In contrast to this, a sharp switching with large coercive field of ~4 T is observed in $Fe_{1/4}TaS_2$; this is typically observed in this material and indicates strong perpendicular anisotropy [1, 2].

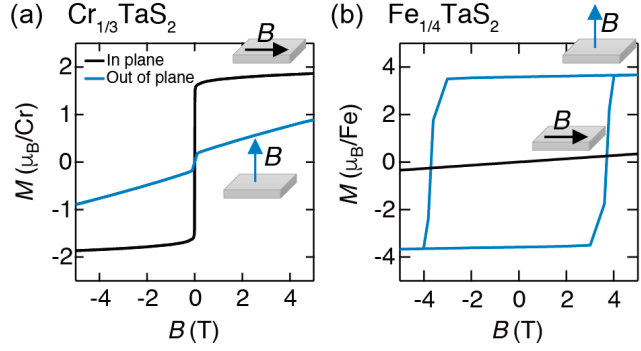

**Figure 2** (a,b) External magnetic field $B$ dependence of magnetization $M$ measured in both in-plane and out-of-plane field directions for (a) $Cr_{1/3}TaS_2$ and (b) $Fe_{1/4}TaS_2$.

We found that $Cr_{1/3}TaS_2$ exhibited preferential exfoliation between the layers upon mechanical exfoliation. Optical micrographs of exfoliated crystals deposited on 300 nm $SiO_2$/Si substrate are shown in Figs. 3(a-d). We note that this is the first reported demonstration of the mechanical exfoliation of this material. The different thickness of the $Cr_{1/3}TaS_2$ exhibits different contrast on the substrate due to the optical interference, and thus can be used to evaluate the thickness of the flake. Exfoliated flakes in the thickness range of nine up to a few hundred nanometres were obtained. Flakes thinner than this were difficult to evaluate because their lateral dimensions are too small. This limits the extension of our study in terms of studying thinner flakes. We think that the flake preparation method must be further improved in future experiments. The AFM images of the sub-hundred-nm-thick exfoliated crystals (Figs. 3(b-d)) are shown in Figs. 3(e-h). The flat surfaces with the root mean square roughness of 0.35-0.50 nm are obtained from the AFM topography image. Such flat surfaces are another piece of evidence that preferential exfoliations of $Cr_{1/3}TaS_2$ occur at the vdW interface. From AFM and optical microscopy characterisations, we did not observe serious degradation of the flakes, such as decomposing in air. This differs from the previously reported layered material ferromagnet $CrI_3$ [20].

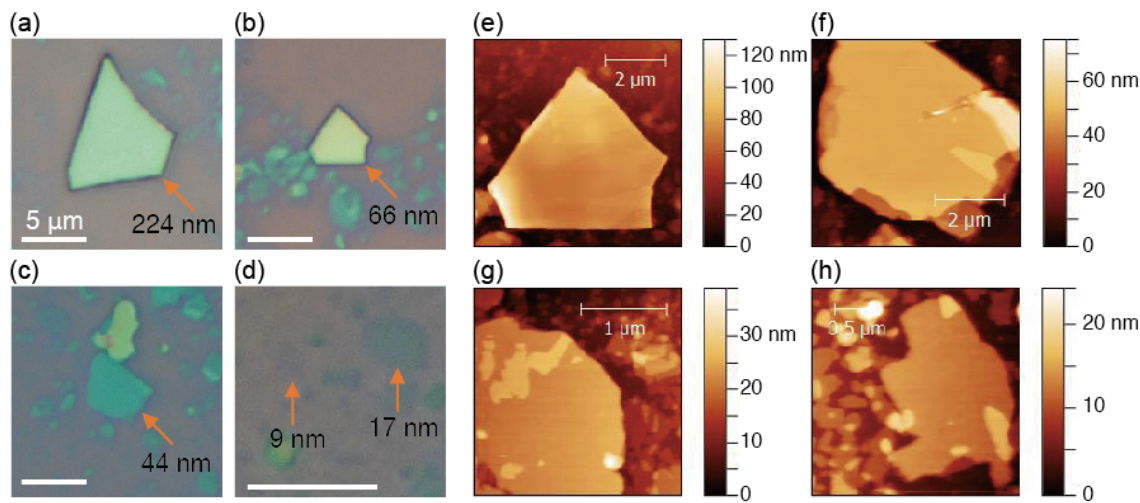

**Figure 3** (a-d) Optical micrographs of the exfoliated $Cr_{1/3}TaS_2$ flakes with different thicknesses. Thicknesses obtained from AFM measurements are indicated. White scale bars in the figures correspond to 5 μm. (e-h) AFM topography images of the exfoliated $Cr_{1/3}TaS_2$ flakes with thicknesses of (e) 66 nm, (f) 44 nm, (g) 17 nm, and (h) 9 nm, respectively.

As we previously showed for the case of $Fe_{1/4}TaS_2$ [12], the intercalated $TaS_2$ oxidizes to form $Ta_2O_5$ in atmosphere; thus, as exfoliation, this material provides the structure of a ferromagnet covered with an insulator. Note that oxidation can take place layer-by-layer owing to the layered crystal structure of the material, forming a smooth oxide layer [29, 30]. We analysed the thickness of oxide on the exfoliated surface of $Cr_{1/3}TaS_2$ and $Fe_{1/4}TaS_2$ flakes with a high-angle annular dark-field (HAADF) STEM image, as shown in Fig. 4. We found pronounced oxidation on the $Fe_{1/4}TaS_2$ surface (~1.2 nm in thickness) and weaker oxidation on the $Cr_{1/3}TaS_2$ surface (less than 0.3-nm thick). Note that since the exfoliated flakes were exposed to air for a few days before the TEM observation, the measured thickness of the oxide should be considered as the upper limit of the native oxide layer present on the surface of tens-of-nanometres-thick $Cr_{1/3}TaS_2$ and $Fe_{1/4}TaS_2$ flakes.

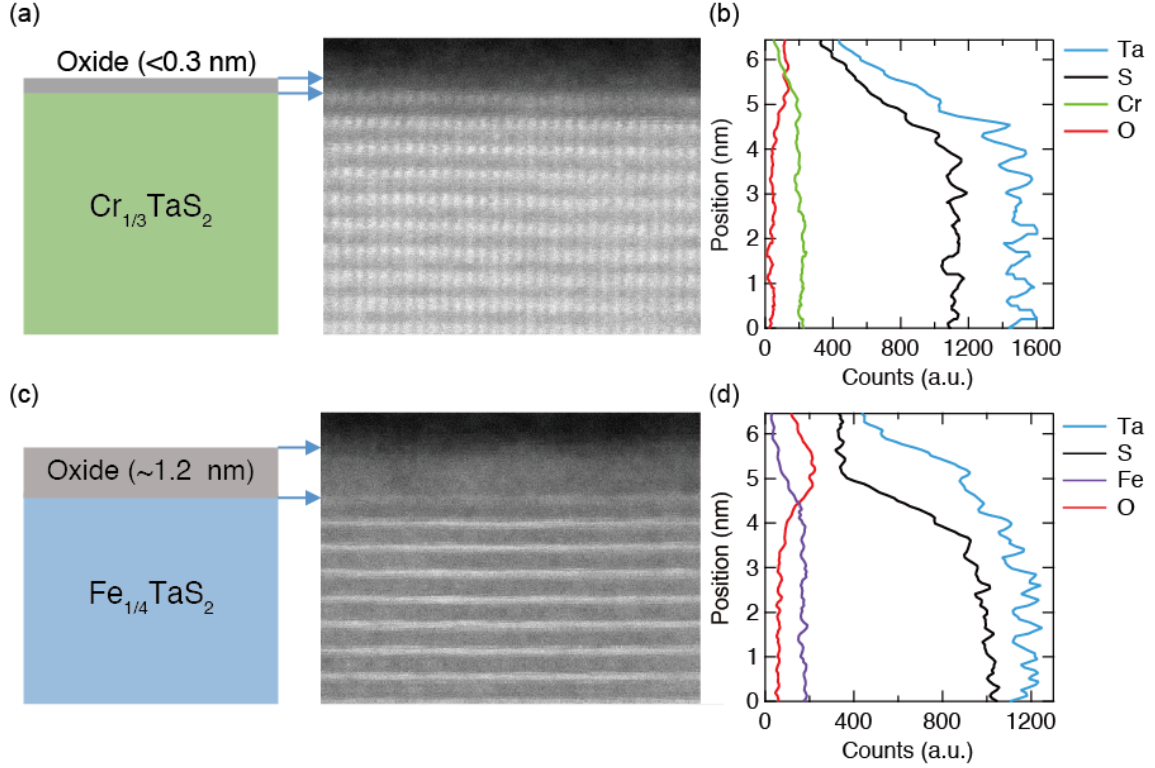

**Figure 4** (a,c) Cross-sectional HAADF STEM image and (b,d) EDX analysis of (a,b) exfoliated $Cr_{1/3}TaS_2$ and (c,d) exfoliated $Fe_{1/4}TaS_2$ surfaces. The schematic illustrations of the crystal structure are also depicted in panels (a) and (c). The depth profile of the EDX signal for Ta, S, Fe, Cr, and O are plotted.

Magnetotransport properties of the exfoliated ~65-nm-thick flake of $Cr_{1/3}TaS_2$ are measured in the device shown in Fig. 5(a) and results are presented in Figs. 5(b)−5(d). For a comparison, the magnetotransport was measured on bulk $Cr_{1/3}TaS_2$; the device micrograph and measurement results are shown in Figs. 5(e)−5(h). The Hall resistance $R_{xy}$ of the $Cr_{1/3}TaS_2$ flake measured at 10 K with an applied magnetic field perpendicular to the sample plane is shown by the black solid line in Fig. 5(b). Here, $R_{xy}$ exhibits a non-linear change with respect to the applied magnetic field, suggesting the contribution of the anomalous Hall effect. With the presence of the anomalous Hall effect, the Hall resistance can be written as $R_{xy} = R_0 B + R_{xy}^A$, where, $R_0$ denotes the ordinal Hall coefficient, $R_{xy}^A = R_s M$ the anomalous Hall term, $R_s$ the anomalous Hall coefficient, and $M$ the magnetization of the film. Since the ordinal Hall coefficient $R_0$ is linear to $B$, we infer this component from the slope at high field region (12 < $B$ < 14.5 T) in Fig. 5(b) and subtract it from the $R_{xy}$ data. The subtracted $R_{xy}$ that is in principle equal to $R_{xy}^{A}$ is plotted as a red solid line in the figure; this displays a clear signature of hard-axis magnetization behaviour. The saturation field $H_k \sim 12$ T shows good coincidence with the value obtained from the $M$-$B$ curve in Fig. 2(a); this value corresponds to the in-plane-anisotropy energy density $K=M_sH_k/2$ of $2 \times 10^5$ J/m$^3$, where $M_s = 42$ mT denotes the saturation magnetization and $H_k$ the anisotropy field. The obtained value of $K$ is larger than the demagnetization energy of $-M_s^2/2\mu_0$, suggesting strong in-plane magnetocrystalline anisotropy, where $\mu_0$ denotes the vacuum permeability. The in-plane anisotropy can be confirmed from magnetic field dependence of $R_{xx}$ measured under in-plane and out-of-plane directions at 10 K as shown in Fig. 5(c). Owing to the AMR contribution, the resistance is lower when $M$ is along the current flow compared to the situation where $M$ is perpendicular to the current. Fig. 5(c) displays a clear signature of AMR and that is consistent with the in-plane anisotropy. These results demonstrate that the exfoliated Cr$_{1/3}$TaS$_2$ exhibits ferromagnetic order at low temperature. The $T_C$ of the flake is determined from the temperature dependence of $R_{xx}$ as shown in Fig. 5(d). The $R_{xx}$ displays significant reduction below $T_C$ due to the suppression of spin disorder scattering [4, 22, 24]. From the data, $T_C \sim 110$ K is determined. We performed $R_{xy}$, AMR, and $R_{xx}$ vs. $T$ measurements on the bulk crystal of Cr$_{1/3}$TaS$_2$ as shown in Figs. 5(f)−5(h); the data between exfoliated and bulk Cr$_{1/3}$TaS$_2$ show good agreement. These findings reveal that the magnetism of Cr$_{1/3}$TaS$_2$ is preserved upon mechanical exfoliation down to tens-of-nanometres-thick flakes.

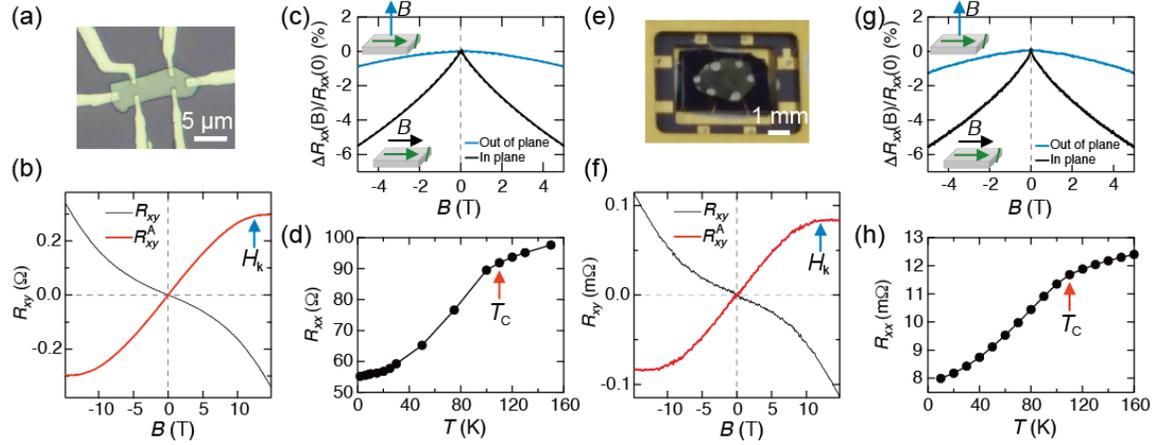

**Figure 5** (a,e) Optical micrograph of (a) one of the exfoliated $Cr_{1/3}TaS_2$ and (e) the bulk $Cr_{1/3}TaS_2$ device. (b,f) Hall resistance $R_{xy}$ measured at 10 K measured in (b) the exfoliated $Cr_{1/3}TaS_2$ flake with a thickness of ~82 nm and (f) the bulk $Cr_{1/3}TaS_2$ device. (c,g) Longitudinal resistance $R_{xx}$ at 10 K with a magnetic field parallel and perpendicular to the plane measured in (c) the exfoliated $Cr_{1/3}TaS_2$ flake with a thickness of ~65 nm and (g) the bulk $Cr_{1/3}TaS_2$. The magnetic-field dependent $R_{xx}(B)$ is normalized to its zero-field value $R_{xx}(0)$. (d,h) The temperature dependence of resistivity measured in (d) the exfoliated $Cr_{1/3}TaS_2$ flake with a thickness of ~65 nm and (h) the bulk $Cr_{1/3}TaS_2$. The measurement was performed by applying ac current $I_{AC}$ = 5 μA for the exfoliated devices, and dc current $I_{DC}$ = 50 mA for the bulk device.

In the $R_{xx}$−$B$ curve measured under the sweep of the small in-plane magnetic field, we observed a clear difference between the exfoliated flake and bulk $Cr_{1/3}TaS_2$ as shown in Figs. 6(a) and 6(b). The hysteresis loop is observed in exfoliated flake (Fig. 6(a)) having two different switching fields $B_1$ and $B_2$. By contrast, no hysteresis is observed in the bulk sample (Fig. 6(b)). Since both the $M$−$B$ curve (Fig. 2(a)) and $R_{xx}$−$B$ of the bulk $Cr_{1/3}TaS_2$ did not show any hysteresis, we think that the observed hysteresis in Fig. 6(a) is due to the consequence of the exfoliation. Appearance of hysteresis in ferromagnetic thin film can be due to the pinning of single or multiple domain walls [31]. In this figure, the domain wall (or walls) created at $B_1$ is pinned within the film and annihilated at $B_2$. This can be confirmed by measuring the angular dependence of $B_1$ and $B_2$. Since an out-of-plane magnetic field cannot drive the domain wall to move, two switching fields only scale with the in-plane components of the magnetic field. $R_{xx}$−$B$ curves were measured for different angles of the external field $\theta$ as shown in Fig. 6(c). From the figure, $B_1$ and $B_2$ are plotted against $\theta$ in Fig. 6(d). Both field values monotonically increase with increasing $\theta$, supporting our speculation that there is significant domain wall pinning in an exfoliated $Cr_{1/3}TaS_2$ flake that changes its switching property. Although the origin of the pinning in the exfoliated $Cr_{1/3}TaS_2$ is unclear at the moment, we infer that it is due to the presence of edge roughness or surface defects in the tens-of-nanometres-thick films.

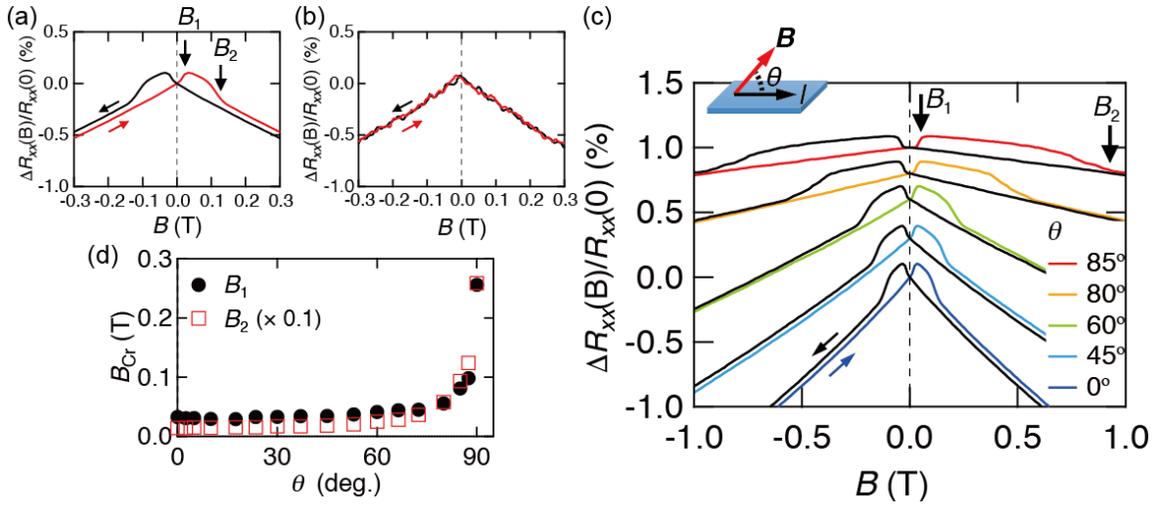

**Figure 6** (a,b) $R_{xx}$–$B$ curves measured in (a) an exfoliated flake of $Cr_{1/3}TaS_2$ and (b) bulk $Cr_{1/3}TaS_2$ crystal. The magnetic field was applied in-plane. The switching fields $B_1$ and $B_2$ are indicated by arrows in the figure. (c) The angle-dependence of $R_{xx}$–$B$ curves measured on the device at $T = 10$ K. The measurement was performed by applying $I_{AC} = 5$ μA. The curve except for $\theta = 0°$ is offset for clarity. Inset: the relationship between the applied current and magnetic field angle $\theta$. (d) Angular dependence of $B_1$ and $B_2$.

Demonstrating the mechanical exfoliation of $Cr_{1/3}TaS_2$ provides the opportunity for vdW heterostructure assembly of this material together with other layered materials. Here, a magnetic tunnel junction built from a heterojunction between $Cr_{1/3}TaS_2$ and $Fe_{1/4}TaS_2$ was fabricated by the dry transfer method. We selected $Fe_{1/4}TaS_2$ as another ferromagnetic electrode because this material had already been demonstrated to be mechanically exfoliatable, and used to fabricate vdW junctions. Micrographs of the exfoliated flakes and fabricated heterojunction device are shown in Fig. 7(a). The oxidized surfaces of $Cr_{1/3}TaS_2$ and $Fe_{1/4}TaS_2$ are connected with van der Waals force during the transfer of the exfoliated flake of $Cr_{1/3}TaS_2$ on the exfoliated $Fe_{1/4}TaS_2$ surface; therefore, the fabricated heterojunction structure is $Fe_{1/4}TaS_2/Ta_2O_5/Cr_{1/3}TaS_2$ as illustrated in Fig. 7(b). The native $Ta_2O_5$ oxide layer with a thickness of approximately 1.5 nm works as a tunnel barrier between the two intercalated ferromagnets and the spin-polarized tunnelling from the $Cr_{1/3}TaS_2$ and $Fe_{1/4}TaS_2$ can be detected by measuring its magnetoresistance (MR). We note that the magnetic properties of the exfoliated tens-of-nanometres-thick $Fe_{1/4}TaS_2$ flakes are similar to its bulk form, as shown in a previous report [12]. The MR measured from the heterojunction composed of $Fe_{1/4}TaS_2$ and $Cr_{1/3}TaS_2$ under the out-of-plane magnetic field ($\theta = 90°$ in Fig. 7(b)) is shown in Fig. 7(c). To detect the junction resistance, the measurement was conducted by four-terminal measurement such that current $I$ is applied between terminals 5 and 2, and then voltage is measured between terminals 4 and 3 (See Fig. 7(a)). We observed clear hysteresis in the MR loop with sharp resistance switching at around ±0.15 T and ±5 T. When the external field was varied from -9 T to +9 T, we observed a resistance increase at +0.15 T and then a continuous increase of resistance between ~0.15 T and ~5 T. Finally, the junction resistance dropped abruptly at ~5 T. The overall junction resistance changes are explained by considering the relative magnetization direction between $Cr_{1/3}TaS_2$ and $Fe_{1/4}TaS_2$. First, comparing Fig. 2(a) with Fig. 7(c), the resistance jump at ~0.15 T is attributed to the partial change of the magnetization in the $Cr_{1/3}TaS_2$ layer. Further, $Cr_{1/3}TaS_2$ undergoes gradual canting along the field direction because the field direction is along its hard axis; this corresponds to the continuous increase of junction resistance from ~0.15 T to ~5 T. Finally, the resistance jump at ~5 T is due to the magnetization reversal of $Fe_{1/4}TaS_2$. Therefore, the data shown in Fig. 7(c) demonstrates the TMR effect in the heterojunction composed of $Fe_{1/4}TaS_2$ and $Cr_{1/3}TaS_2$. To support this, we found that the junction resistance shown in Fig. 7(c) is significantly higher than the bulk resistance of either $Cr_{1/3}TaS_2$ or $Fe_{1/4}TaS_2$ due to the presence of a tunnel barrier at the junction.

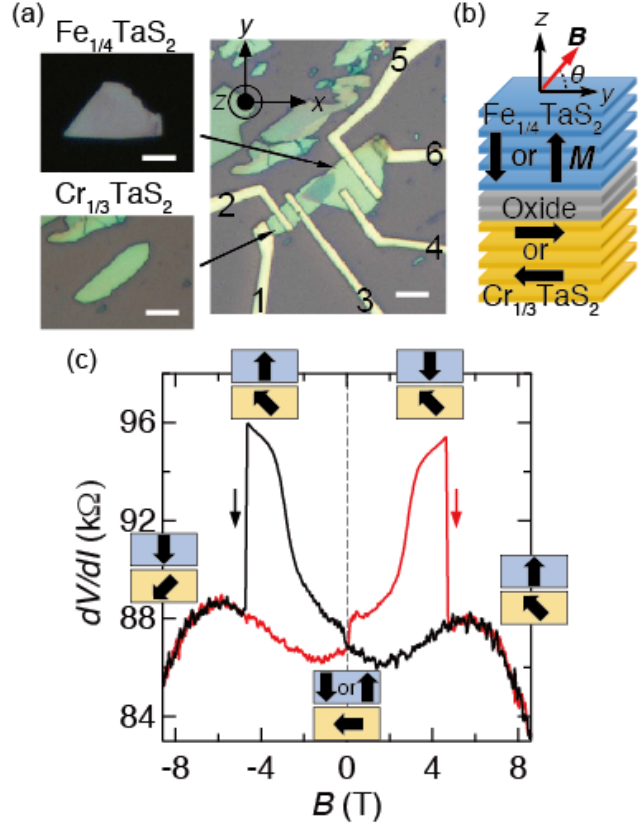

**Figure 7** (a) Schematic illustration and optical micrographs of exfoliated $Cr_{1/3}TaS_2$, $Fe_{1/4}TaS_2$, and heterojunction composed of the $Fe_{1/4}TaS_2$ and $Cr_{1/3}TaS_2$ device. The $Cr_{1/3}TaS_2$ flake was deposited on $SiO_2$/Si substrate and the $Fe_{1/4}TaS_2$ flake was deposited on PDMS. The thickness of $Cr_{1/3}TaS_2$ and $Fe_{1/4}TaS_2$ was ~50 nm and ~40 nm, respectively. The white bar in the figure represents a scale of 5 μm. (b) Illustration of the structure of the fabricated heterojunction. The arrows indicate the preferential magnetization direction $M$ of the $Cr_{1/3}TaS_2$ and $Fe_{1/4}TaS_2$ flakes. The magnetic field angle $\theta$ is also shown. (c) Magnetic field dependence of differential resistance $dV/dI$ at $T = 8$ K of the heterojunction composed of $Fe_{1/4}TaS_2$ and $Cr_{1/3}TaS_2$ measured at the $I_{AC} = 10$ nA and $I_{DC} = +50$ nA. The magnetic field was applied perpendicular to the plane ($\theta = 90°$).

To further investigate the magnetization reversal of each electrode, the MR was measured for different angles $\theta$ of the external magnetic field, and the results are shown in Figs. 8(a) and 8(b) (see Fig. 7(b) for the definition of angle $\theta$). We observed MR loop hysteresis over a wide range of $\theta$ values. The two switching fields $B_3$ and $B_4$ indicated by arrows in Figs. 8(a) and 8(b) change with respect to $\theta$. The values of $B_3$ and $B_4$ against $\theta$ are shown in Figs. 8(c) and 8(d), respectively. Here, $B_3$ tends to decrease from $\theta = 0°$ to $90°$, preferring switching in the out-of-plane direction. By contrast, the switching field $B_4$ tends to increase in the same range of $\theta$. The $\theta$ dependence on $B_4$ shows good agreement with that of $Cr_{1/3}TaS_2$ in Fig. 6(d). These results support our speculation that $B_3$ is caused by the magnetization reversal of $Fe_{1/4}TaS_2$ and $B_4$ is caused by the partial magnetization canting of the $Cr_{1/3}TaS_2$ layer. In fact, the $B_3$ vs. $\theta$ curve can be fitted with the relationship $B_3 = B_\perp/\sin\theta$ (Fig. 8(c)) with $B_\perp = 3.5$ T; this relationship is valid for films with strong perpendicular anisotropy. Owing to strong perpendicular magnetic anisotropy, the magnetization of the $Fe_{1/4}TaS_2$ remains along the out-of-plane direction for most of the external magnetic field angles. In comparison, the $Cr_{1/3}TaS_2$ magnetization follows the external magnetic field direction because it has smaller anisotropy. Therefore, we observed a hysteresis loop for a wide range of $\theta$. These results clearly indicate that the MR behaviour in heterojunction composed of $Fe_{1/4}TaS_2$ and $Cr_{1/3}TaS_2$ is determined by the relative magnetization alignment between the two ferromagnetic layers.

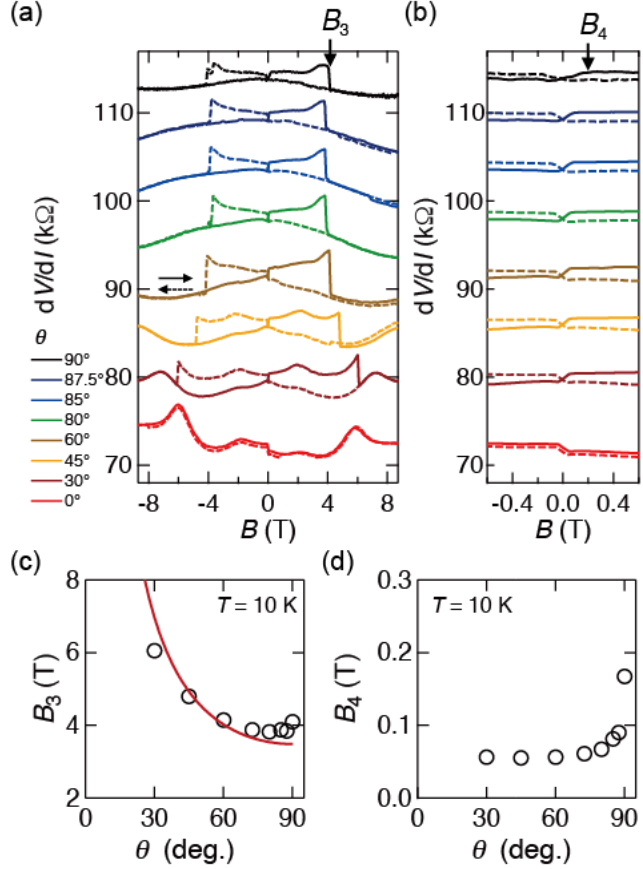

**Figure 8** (a,b) Angular-dependence of MR of the heterojunction composed of $Fe_{1/4}TaS_2$ and $Cr_{1/3}TaS_2$ at $T = 10$ K. The measurement was performed by applying $I_{AC} = 10$ nA and $I_{DC} = +100$ nA. Both (a) the full range of the magnetic field and (b) the magnification around low magnetic field regions are presented. The curve except for $\theta = 0°$ is offset for clarity. The positions of the switching fields $B_3$ and $B_4$ are also indicated. Arrows indicate the sweep directions of the external magnetic field. (c,d) Angular dependence of (c) $B_3$ and (d) $B_4$. The solid line in Fig. 5(c) plots the relationship $B_3 = B_\perp/\sin\theta$.

Next, we measured the bias current $I_{DC}$ dependence of magnetoresistance (MR), as shown in Fig. 9(a). We measured the amplitude of resistance jump at $B \sim 4.0$ T as $\Delta R$. Then, MR ratio = $[\Delta R/R(B = 4.5$ T$)]\times 100$. The bias dependence of the MR ratio is shown in Fig. 9(b). The MR ratio decreases with increasing $I_{DC}$, suggesting that it is a TMR effect. Tunnelling transport in the junction is also confirmed from the change of junction resistance $dV/dI$ with respect to $I_{DC}$ at 10 K, as shown in Fig. 9(c); $dV/dI$ also decreases with increasing bias. Noticeably, the maximum

MR ratio of 13% is significantly higher than that in previous experiments with a junction between the different flakes of $Fe_{1/4}TaS_2$ (~7%) fabricated using the same method [12]. This suggests a larger tunnel spin polarization when using $Cr_{1/3}TaS_2$.

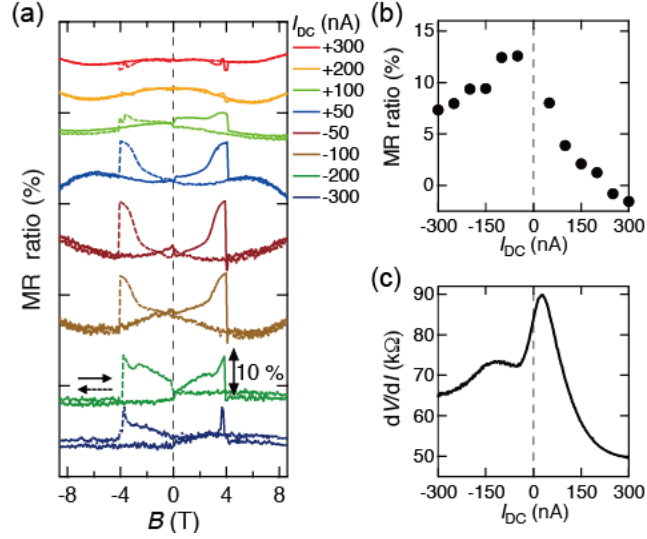

**Figure 9** (a) The dc current bias-dependence of MR curves measured on the heterojunction composed of $Cr_{1/3}TaS_2$ and $Fe_{1/4}TaS_2$ at $T = 10$ K. Curves are offset for clarity. (b) The dc current bias-dependence of the MR ratio at $T = 10$ K. (c) The dc current bias-dependence of $dV/dI$ measured at $T = 10$ K and zero applied magnetic field.

Further, the temperature dependence of the MR loops in the heterojunction composed of $Cr_{1/3}TaS_2$ and $Fe_{1/4}TaS_2$ is shown in Fig. 10(a), and the MR ratio for $I_{DC} = -50$ nA is plotted against temperature, as shown in Fig. 10(b). The MR decreases with increasing temperature and disappears above 35 K. In comparison, the junction resistance $dV/dI$ at $I_{DC} = 0$ A and zero magnetic field is measured at a different temperature, as shown in the Fig. 10(c). It is clearly seen that the junction resistance becomes weakly temperature-dependent below 20~30 K, indicating tunnelling transport. MR is observed below this temperature range. Elevating the temperature introduces additional conduction paths because of the thermal broadening of the Fermi level, and this significantly reduces junction resistance. The temperature dependence of a resistance in the heterojunction composed of Fe$_{1/4}$TaS$_2$ and Cr$_{1/3}$TaS$_2$ shown in the Fig. 10(c) is fitted with the model based on the thermal broadening contribution model [32]. Within this model, the temperature dependence of zero-bias resistance $dV(T)/dI$ data is fitted with the function $dV(T)/dI = R_0 \sin CT / CT$, where $R_0$ is the zero-bias resistance at 10 K, $C = 2\pi^2 k_B d \sqrt{2m^*} / \sqrt{h^2 \varphi}$, $k_B$ is the Boltzmann constant, $d$ the thickness of the tunnel barrier, $m^*$ the effective mass inside the tunnel barrier, $h$ the Planck constant, and $\varphi$ the tunnel barrier height. The result of this fitting is plotted as a solid line in Fig. 10(c). With oxide thickness of ~1.5 nm determined from TEM and tunnelling effective mass $m^* = m_0$ ($m_0$ denotes electron mass), we demonstrate reasonably good agreement between the experimental and fitted data with a tunnel barrier height of $\varphi \sim 4.4$ meV. Both bias and temperature dependences of MR are clear evidence of the TMR effect demonstrated in the heterojunction composed of Cr$_{1/3}$TaS$_2$ and Fe$_{1/4}$TaS$_2$ with a native T$_2$O$_5$ oxide tunnel barrier in between. The determined tunnel barrier height is lower than a typical value obtained for Ta$_2$O$_5$ [33], suggesting the low quality of native oxide as a tunnel barrier. We expect that by using other high-quality 2D materials as tunnel barriers such as h-BN and MoS$_2$, rather native Ta$_2$O$_5$, we can increase the MR ratio further. Indeed, these materials can be integrated with a vdW heterostructure assembly technique [34-37].

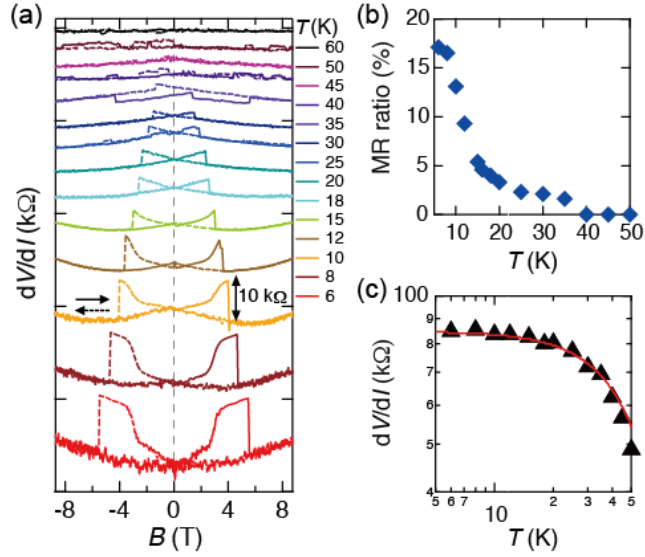

**Figure 10** (a) Temperature-dependence of MR curves measured under $I_{DC}$ = -50 nA. The curves are offset for clarity. (b) Temperature-dependence of the MR ratio at $I_{DC}$ = -50 nA. (c) Temperature-dependence of $dV/dI$ measured at $I_{DC}$ = 0 A and zero magnetic field. The solid line is the fitting result based on the thermal broadening contribution model, as explained in the main text.

## 4. Conclusion.

We demonstrated the mechanical exfoliation of magnetic-atom-intercalated transition metal dichalcogenide $Cr_{1/3}TaS_2$ and the vdW heterostructure assembly of exfoliated flakes using the dry transfer method. The exfoliated $Cr_{1/3}TaS_2$ exhibits ferromagnetic ordering down to tens-of-nanometres-thick flakes, and the fabricated vdW heterojunction composed of $Cr_{1/3}TaS_2$ and $Fe_{1/4}TaS_2$ with a native $T_2O_5$ oxide tunnel barrier in between exhibits the TMR effect. These findings demonstrate the possibility of using magnetic-atom-intercalated TMD materials in the field of 2D-material spintronics.

**Acknowledgements**

This work was supported by CREST, Japan Science and Technology Agency (JST) Grant Number JPMJCR15F3, JSPS KAKENHI Grant Numbers JP25107003, JP25107004, JP26248061, JP15H01010, JP15K17433, JP16H00982, and Murata Science Foundation.